\documentclass[prb,superscriptaddress,twocolumn,showpacs,amsmath,amssymb]
{revtex4}
\usepackage{graphicx}
\usepackage{dcolumn}
\usepackage{bm}
\usepackage{txfonts}
\usepackage{mathrsfs}
\usepackage{feynmf}
\usepackage{comment}
\def\bk{{\mathbf k}}

\def\bq{{\mathbf q}}

\def\b0{{\mathbf 0}}

\def\up{\uparrow}
\def\down{\downarrow}

\def\alf{\alpha}
\def\eps{\epsilon}

\def\Gam{\Gamma}
\def\lam{\lambda}
\def\Lam{\Lambda}

\def\sg{\sigma}

\def\psib{\bar\psi}

\usepackage[]{psfrag}

\psfrag{pp_etas}{$\eta_{f},\eta_{b}$}
\psfrag{p_e_b}{$\eta_{b}$}
\psfrag{p_e_f}{$\eta_{f}$}
\psfrag{Z_b}{$Z_{b}$}
\psfrag{Z_f}{$Z_{f}$}
\psfrag{pp_g}{$\tilde{g}$}
\psfrag{pp_mass}{$\tilde{m}^{2}_{b}$}
\psfrag{pp_u}{$\tilde{u}$}
\psfrag{log_chi}{$\log\left[\xi^{-2}\right]$}
\psfrag{log_mb}{$\log\left[m^{2}_{b}-m^{2}_{b,\text{crit}}\right]$}

\psfrag{d_del}{$\delta-\delta_{\text{c}}$}
\psfrag{tyldu}{$\tilde{u}$}
\psfrag{tyldel}{$\tilde{\delta}$}
\psfrag{delta}{$\delta$}
\psfrag{chi_inv}{$\chi^{-1}$}
\psfrag{L}{$\Lambda$}
\psfrag{w}{$\xi_{\mathbf{k}}$}

\begin{document}

\title{Anomalous scaling of fermions and order parameter fluctuations 
at quantum criticality}
\author{P. Strack}
\email{p.strack@fkf.mpg.de}
%
\author{S. Takei}
%
\author{W. Metzner}
\affiliation{Max-Planck-Institute for Solid State Research,
 Heisenbergstr.\ 1, D-70569 Stuttgart, Germany} 
\date{\today}
\begin{abstract}
We analyze the quantum phase transition between a semimetal and a
superfluid in a model of attractively interacting fermions with a
linear dispersion.
The quantum critical properties of this model cannot be treated
by the Hertz-Millis approach since integrating out the fermions
leads to a singular Landau-Ginzburg order parameter functional.
We therefore derive and solve coupled renormalization group 
equations for the fermionic degrees of freedom and the bosonic 
order parameter fluctuations.
In two spatial dimensions, fermions and bosons 
acquire anomalous scaling dimensions at the quantum critical point, 
associated with non-Fermi liquid behavior and non-Gaussian order 
parameter fluctuations.
\end{abstract}

\maketitle


\section{Introduction}

As noticed already more than 30 years ago by Hertz,\cite{hertz76}
quantum phase transitions in correlated fermion systems
may be described in terms of the order parameter field alone 
when one integrates out fermions from the path integral in one 
stroke and subsequently deals with an exclusively bosonic theory. 
The resultant effective action is expanded in powers of the bosonic 
field $\phi$ and often truncated after the $\phi^{4}$-term. 
Although an analysis in terms of $\phi^{4}$-type theories seems 
mundane, the presence of \emph{two} relevant energy scales, one 
given by temperature and the other given by a non-thermal control 
parameter acting as a mass term in the bosonic propagator, gives 
rise to a rich finite temperature phase diagram.\cite{millis93}

The Hertz-Millis approach relies on integrating out fermions first.
In general, however, the fermions are gapless and consequently 
may lead to singular coefficients in the effective bosonic action.
\cite{belitz05,loehneysen07}
In such cases, it is advisable to keep the fermions in the theory 
and treat them on equal footing with the bosons.
Various coupled Fermi-Bose systems exhibiting quantum criticality 
have been analyzed previously.
\cite{altshuler94,vojta00,abanov00,abanov03,rech06,kaul08,huh08}
The complicated interplay of two singular propagators promotes a
controlled treatment to a formidable task.

A suitable tool to cope with the interplay of fermionic and bosonic 
fluctuations is the functional renormalization group (RG) formulated 
for fermionic and bosonic fields.
\cite{berges02,baier04,schuetz05,strack08} 
\begin{figure}
\vspace*{5mm}
\hspace*{4mm}
\includegraphics*[width=75mm,angle=0]{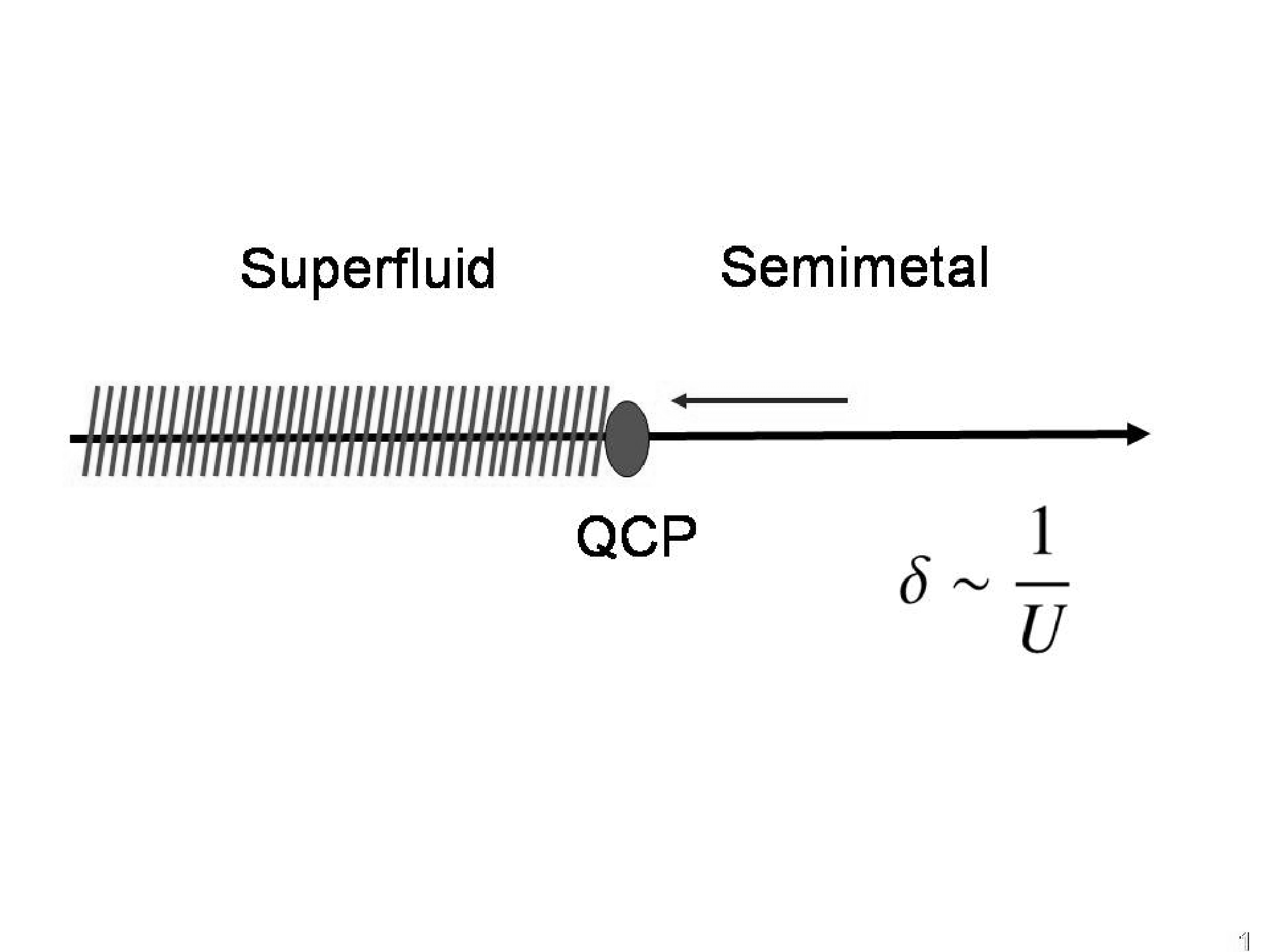}
\vspace*{-17mm}
\caption{Schematic phase diagram of the attractive Dirac cone model 
at zero temperature. The quantum critical point (QCP) at a critical interaction 
strength $U_{c}$ separates the semimetal from the superfluid. 
In this paper, we approach the QCP from the semimetallic 
phase as indicated by the arrow.}
\label{fig:phase_toy}
\end{figure}
The mutual feedback of gapless fermions coupled to massless bosons
has been studied already with functional flow equations for quantum 
electrodynamics,\cite{gies04} non-abelian gauge theories,
\cite{pawlowski04} and the Gross-Neveu model.\cite{rosa01}

The purpose of the present paper is to show how one can -- within 
the coupled Fermi-Bose RG framework -- obtain non-Fermi liquid 
properties of quantum critical fermion systems for which the 
Hertz-Millis approach is not applicable.
To this end, we study a ``Dirac cone'' model of attractively 
interacting fermions with a linear dispersion, 
which exhibits a quantum phase transition from a semimetal to a 
superfluid at a finite interaction strength as shown in 
Fig.~\ref{fig:phase_toy}. 
The coupled fermion-boson flow equations are physically transparent 
and a relatively simple truncation seems to capture the essential 
renormalizations. 
As a central result, the fermion and order parameter propagators
develop anomalous power law dependences on frequency and momentum 
at the quantum critical point (QCP) in dimensions $d<3$.

In Sec.~II, we introduce the attractive Dirac cone model 
and the associated Fermi-Bose action. Subsequently,
in Sec.~III, we show that a mean-field treatment of this model 
leads to a semimetal-to-superfluid quantum phase transition 
at a critical interaction strength $U_c$. 
In Sec.~IV, the RG method, truncation and flow equations are
presented. Results for the quantum critical behavior follow in 
Sec.~V.
We finally summarize and conclude in Sec.~VI.


\section{Dirac cone model}

\begin{figure}
\begin{center}
\vspace*{3mm}
\includegraphics*[width=56mm]{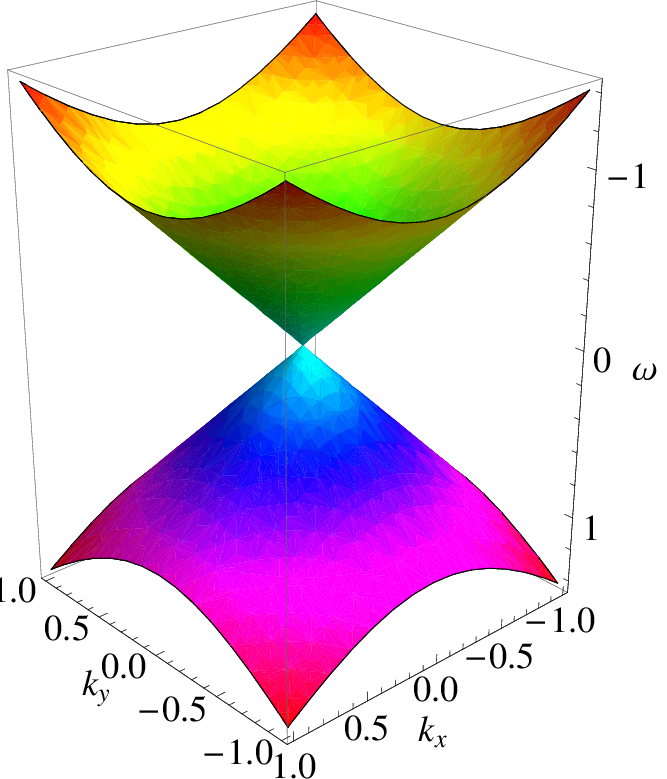}
\caption{(Color online)
``Dirac cone'' dispersion in two dimensions.}
\label{fig:cone}
\end{center}
\end{figure}

We consider an interacting fermion system with the bare action 
\begin{eqnarray}
  \Gam_0[\psi,\psib]
  &=& - \int_{k\alf\sg}
  \psib_{k\alf\sg} (ik_{0}-\xi_{\bk\alf}) \, \psi_{k\alf\sg}
 \nonumber\\
  &+& U \! \int_{k\alf} \int_{k'\alf'} \! \int_q 
  \psib_{-k,\alf\down} \psib_{k+q,\alf\up}
  \psi_{k'+q,\alf'\up} \psi_{-k',\alf'\down} \, .
  \label{eq:bare_action}
\end{eqnarray}
The variables $k = (k_0,\bk)$ and $q = (q_0,\bq)$ collect Matsubara 
energies and momenta, and we use the short-hand notation
$\int_k =
 \int_{-\infty}^{\infty} \frac{d k_{0}}{2\pi}
 \int \frac{d^d \bk}{(2\pi)^d} \,$
for momentum and energy integrals;
$\int_{k\alf}$ includes also the sum over the band index $\alf=\pm 1$
and $\int_{k\alf\sg}$ includes also the spin sum over 
$\sg = \, \up,\down \,$.
Our analysis is restricted to zero temperature such that the
Matsubara energies are continuous variables.
The dispersion of the fermions is given by the ``Dirac cone''
\begin{eqnarray}
\xi_{\bk\alf} = \alf v_f |\bk| \; ,
\label{eq:dispersion}
\end{eqnarray}
corresponding to massless particles with positive ($\alf=1$) 
and negative ($\alf=-1$) energy.
The chemical potential is chosen as $\mu = 0$, such that in
the absence of interactions states with negative energy are 
filled, while states with positive energy are empty. The
Fermi surface thus consists only of one point, the ``Dirac 
point'' at $\bk=\b0$, where the two cones 
in Fig.~\ref{fig:cone} intersect.
Momentum integrations are cut off in the ultraviolet by the 
condition $|\xi_{\bk\alf}| < \Lambda_0$.

The kinetic energy in Eq.~(\ref{eq:bare_action}) is a simplified 
version of the dispersion for electrons moving on a honeycomb lattice 
as in graphene, where the momentum dependence is entangled with a
pseudospin degree of freedom related to the two-atom structure
of the unit cell.\cite{castroneto09}

The two-particle interaction in Eq.~(\ref{eq:bare_action})
is parametrized by a momentum independent coupling constant $U$ 
and corresponds to a local interaction in real space. 
Note that the kinetic energy and the interaction are both 
diagonal in the spin indices.
Although our model is reminiscent of the Gross-Neveu model,
\cite{gross74} it is not equivalent to it. In particular, 
for the Gross-Neveu model there is no choice of a spinor basis 
in which the kinetic and potential energies are both 
spin-diagonal.
The Gross-Neveu model is Lorentz-invariant while our model is not.

For attractive interactions, the coupling constant $U$ is negative 
and drives spin singlet pairing associated with spontaneous breaking 
of the global $U(1)$ gauge symmetry. 
Therefore, we decouple the interaction in the s-wave
spin-singlet pairing channel by introducing a complex bosonic
Hubbard-Stratonovich field $\phi$ conjugate to the bilinear
composite of fermionic fields
$U \int_{k\alf} \psi_{k+q,\alf\up} \psi_{-k,\alf\down} \,$.
\cite{popov87}
This yields a functional integral over $\psi$, $\psib$ and $\phi$
with the new bare action
\begin{eqnarray}
 \Gam_0[\psi,\psib,\phi]
  &=& - \int_{k\alf\sg} 
  \psib_{k\alf\sg} (ik_0 - \xi_{\bk\alf}) \, \psi_{k\alf\sg}
  - \int_q \phi^*_q \frac{1}{U} \, \phi_q
  \nonumber\\
   &+& \! \int_{k\alf} \int_q \left(
  \psib_{-k,\alf\down} \psib_{k+q,\alf\up} \,
  \phi_q +
  \psi_{k+q,\alf\up} \psi_{-k,\alf\down} \,
  \phi^*_q \right) \, , \nonumber\\ 
  \label{eq:dirac_finalmodel}
\end{eqnarray}
where $\phi^*$ is the complex conjugate of $\phi$, while $\psi$
and $\psib$ are algebraically independent Grassmann variables. 
The boson mass $\delta = - 1/U$ plays the role of the control
parameter for the quantum phase transition.

%


\section{Mean-field theory}

Neglecting bosonic fluctuations by replacing $\phi_{q}$ with its 
expectation value $\phi_{q=0}$, the saddle-point approximation
solves the functional integral of Eq.~(\ref{eq:dirac_finalmodel}) 
and leads to the standard BCS gap equation\cite{popov87}
\begin{eqnarray}
 \phi_{0}= 
 -U \int_{k\alf} 
 \frac{\phi_{0}}{k_{0}^{2}+\xi^{2}_{\bk\alf}+\phi_{0}^{2}}
 \; .
\label{eq:bcs_gap}
\end{eqnarray}
The critical interaction strength $U_{c}$, beyond which the gap 
equation has a solution with non-zero $\phi_{0}$, is given by
\begin{eqnarray}
 \frac{1}{U_{c}} =  
 - \int \frac{d k_{0}}{2\pi}\int^{\Lambda_{0}}_{-\Lambda_{0}} 
 d\xi \, N(\xi) \, \frac{1}{k_{0}^{2}+\xi^{2}} \; ,
\label{eq:thouless}
\end{eqnarray}
where the density of states has the form
\begin{eqnarray}
 N(\xi) = \frac{K_d}{v_f^d} \, |\xi|^{d-1} \;,
\end{eqnarray}
with $K_{d}$ being defined by $\int\frac{d^{d}k}{\left(2\pi\right)^{d}}
= K_{d} \int d|\bk|\, |\bk|^{d-1}$. 
Since the density of states vanishes at the Fermi level (for $d>1$), 
the system is stable against pairing for a weak attraction. 
Instead, a finite attraction beyond a certain threshold is necessary to 
cause superfluidity even at zero temperature. 
Performing the integrations in Eq.~(\ref{eq:thouless}), we obtain the 
mean-field position of the quantum phase transition in 
Fig.~\ref{fig:phase_toy}. 
In two dimensions, we have:
\begin{eqnarray}
 \delta_c^{\text{MFT}} = - \frac{1}{U_c^{\text{MFT}}} = 
 \frac{\Lambda_{0}}{2\pi v_f^2} \;.
\label{eq:Uc_mft}
\end{eqnarray}
Upon setting $v_f = \Lambda_0 = 1$, the numerical value for the 
mean-field control parameter value is $\delta_c^{\text{MFT}} = 0.159$. 

A similar quantum phase transition from a semimetal to a superfluid
has been discussed previously for attractively interacting electrons
on a honeycomb lattice, in the context of ultracold atoms \cite{zhao06}
and graphene.\cite{uchoa07}

In the following, we take fluctuations into account by performing
a renormalization group analysis which enables us to compute 
the quantum critical 
exponents at and in the vicinity of the QCP.


\section{RG Method}
\label{sec:pp_method}

We derive flow equations for the scale-dependent effective action
$\Gamma^{\Lambda}[\psi,\bar{\psi},\phi]$
within the functional RG framework for fermionic and
bosonic degrees of freedom in its one-particle irreducible 
representation.\cite{berges02,baier04,schuetz05,strack08} 
Starting from the bare fermion-boson action
$\Gamma^{\Lambda=\Lambda_0}[\psi,\bar{\psi},\phi] =
 \Gamma_0[\psi,\bar{\psi},\phi]$
in Eq.~(\ref{eq:dirac_finalmodel}), fermionic and bosonic 
fluctuations are integrated \emph{simultaneously}, proceeding
from higher to lower scales as parametrized by the continuous 
flow parameter $\Lambda$. 
In the infrared limit $\Lambda \rightarrow 0$, the fully
renormalized effective action 
$\Gamma^{\Lambda \rightarrow 0}[\psi,\bar{\psi},\phi]$ is 
obtained.

The flow of $\Gamma^{\Lambda}$ is governed by the exact functional 
flow equation \cite{berges02,baier04,schuetz05,strack08}
\begin{eqnarray}
\frac{d}{d\Lam} \Gam^{\Lam}[\psi,\psib,\phi] =
 {\rm Str} \, \frac{\partial_{\Lam} R^{\Lam}}
 {\Gam^{(2) \, \Lam}[\psi,\psib,\phi] + R^{\Lam}} \;,
\label{eq:pp_exact}
\end{eqnarray}
where $\Gam^{(2) \, \Lam}$ denotes the second functional derivative 
with respect to the fields and $R^{\Lam}$ is the infrared regulator
(to be specified below). The supertrace (Str) traces over 
all indices, with an additional minus sign for fermionic contractions. 

When evolving $\Gam^{\Lam}$, infinitely many terms involving 
fermionic and/or bosonic fields with possibly complicated dependences 
on frequencies and momenta are generated, necessitating a truncation of 
the effective action.

\subsection{Truncation}
\label{subsec:toy_trunc}

We now explain how the effective action is truncated with the objective 
to capture the essential renormalization effects.

\subsubsection{Fermion propagator}

To account for the renormalization of the fermionic single-particle
properties by order parameter fluctuations, the quadratic fermionic 
term in the action is multiplied by a field renormalization factor
$Z_f$,
\begin{equation}
 \Gam_{\psib\psi} = - \int_{k\alf\sg}
 \psib_{k\alf\sg} Z_f(ik_0 - \xi_{\bk\alf}) \,
 \psi_{k\alf\sg} \; ,
\label{eq:toy_Gam_psi_psi}
\end{equation}
yielding the fermion propagator
\begin{equation}
 G_{f\alf}(k) = 
 - \langle \psi_{k\alf\sg}\bar{\psi}_{k\alf\sg} \rangle =
 \frac{Z_f^{-1}}{ik_0 - \xi_{\bk\alf}} \; .
\label{eq:pp_green_f}
\end{equation}
A diverging $Z_f$ suppresses the quasi-particle weight to zero,
leading to non-Fermi liquid behavior.
The Fermi velocity is not renormalized separately but kept fixed,
since for the linear dispersion relation, Eq.~(\ref{eq:dispersion}),
the renormalization of the term proportional to $k_0$ and the
one proportional to $\xi_{\bk\alf}$ in $\Gamma_{\psib\psi}$ differ
only by a finite factor of order one.
The initial condition for $Z_f$ is $Z_f=1$. 

\subsubsection{Boson propagator}

The bosonic quadratic part of the bare action, 
Eq.~(\ref{eq:dirac_finalmodel}), consists only of a local mass term.
Integrating out fluctuations, a momentum and frequency dependence 
is generated, in particular by fermionic contributions involving
the fermionic particle-particle bubble. For small momenta and
frequencies this dependence is quadratic, leading to the following 
ansatz for the bosonic quadratic part of the action:
\begin{eqnarray}
 \Gam_{\phi^\ast\phi}=
 \int_{q} \phi^{\ast}_q \, 
 \left[ Z_b \left( q_0^{2} + \bq^2 \right) + \delta \right]
 \phi_q
\;.
\end{eqnarray}
Note that there is no linear term in frequency here, as we 
consider the half-filled band and the usual (imaginary) linear 
frequency part of the particle-particle bubble vanishes exactly 
due to particle-hole symmetry. 
The parameter $\delta$ controls the distance to the quantum phase 
transition and is also renormalized by fluctuations. 
If the initial value of $\delta$ is tuned so that 
$\delta \rightarrow 0$ for vanishing cutoff $\Lam \rightarrow 0$, 
we are at the quantum critical point. 
The boson propagator, parametrized by two flowing parameters,
$Z_b$ and $\delta$, reads
\begin{eqnarray}
G_{b}(q) &=& - \langle\phi_{q}\phi^{\ast}_q \rangle =
 - \frac{1}{Z_b \left(q_0^2 + \bq^2 \right) + \delta} \; .
\label{eq:pp_green_b}
\end{eqnarray}
The initial condition for the control parameter is 
$\delta = - 1/U$.
The initial condition for $Z_b$ corresponding to the bare action
in Eq.~(\ref{eq:dirac_finalmodel}) is $Z_b = 0$. 
However, starting the flow with $Z_b = 0$ leads to very large transient
anomalous dimensions at the initial stage of the flow (for $\Lam$
near $\Lam_0$), which complicates the analysis in a (high energy) 
regime which is physically not interesting. The low energy flow
(small $\Lam \to 0$) does not depend on the initial value of $Z_b$.
We therefore add a term $\int \phi_q^* (q_0^2 + \bq^2) \, \phi_q$ 
to the bare action, corresponding to an initial value $Z_b = 1$.
This term regularizes our model by suppressing the interaction for 
large momentum and energy transfers.

\subsubsection{Order parameter self interaction}

\begin{figure}[b]
\begin{fmffile}{fermion_ring_10}
\begin{eqnarray}
u_{n}&:&\hspace{2mm}
\parbox{30mm}{\unitlength=1mm\fmfframe(2,2)(1,1){
\begin{fmfgraph*}(30,20)
     \fmfpen{thin}
     \fmfsurroundn{e}{10}
     \begin{fmffor}{n}{1}{1}{10}
       \fmf{dbl_wiggly}{e [n],i [n]}
     \end{fmffor}
     \fmfcyclen{plain,tension=10/8}{i}{10}
   \end{fmfgraph*}}}
\nonumber
\end{eqnarray}
\end{fmffile}
\caption{Fermion loop which generates the bosonic 2n-point
 vertex, here for $n=5$.} 
\label{fig:fermion_ring} 
\end{figure}
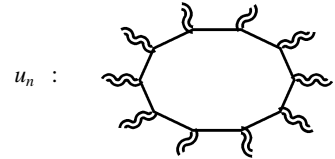

In the Hertz-Millis theory of quantum criticality 
\cite{hertz76,millis93} one expands the order parameter
self interaction in (even) powers of the bosonic fields:
\begin{eqnarray}
 \Gamma_{\text{I}} \left[\phi\right]= \sum_{n \geq 2} u_{2n}
 \int \left(\phi^{\ast}\phi\right)^{n}\;,
\end{eqnarray}
where the coefficients $u_{2n}$ are generated by fermion loops with 
$2n$ fermion propagators and $2n$ external bosonic legs as shown in 
Fig.~\ref{fig:fermion_ring}.
At finite temperatures or when the fermions are gapped, the integrals
corresponding to these loops are finite. 
With the bare, gapless fermion propagator at zero temperature,
however, the loops diverge for vanishing external momenta:
\begin{eqnarray}
 u_{2n} \sim \int d^{d+1}k \, \frac{1}{(k_0^2 + \xi_{\bk}^2)^n} 
 \sim \frac{1}{\Lam^{2n-(d+1)}} \,\,,
\end{eqnarray}
where $\Lam$ is an infrared cutoff introduced here merely to specify 
the degree of the divergence.
In particular, the $\phi^{4}$-interaction diverges as 
$u_{4} \sim 1/\Lam$ in two dimensions.
For the present case of fermionic loops with pairing vertices,
there are no cancellations which reduce the divergence below the level
of simple power counting. This is in stark contrast to the situation
for instabilities driven by forward scattering, where oscillating 
integrands and cancellations between permutations of external legs 
render the loops finite in the low energy limit.\cite{neumayr98,kopper01}

The divergence of the fermion loops makes the Hertz-Millis approach 
of expanding the effective action in powers of the ordering field 
alone inapplicable. Besides issues related to the increasing degree
of divergence of higher order interactions, already at the quartic
level one would face the problem to parametrize the complicated
singular momentum dependence of the interaction generated by the
fermionic loop.
Integrating the fermions along with the bosons in a coupled RG flow,
the vertices remain finite and can be parametrized by a momentum
independent running coupling as usual.
In particular, the $\phi^{4}$-term, which is generated by fermionic
fluctuations, but then also influenced by bosonic fluctuations,
can be written as
\begin{equation}
 \Gam_{|\phi|^4} = \frac{u}{8} \int_{q,q',p}
 \phi^*_{q+p} \phi^*_{q'-p} \phi_{q'} \phi_q \;.
\end{equation}
The initial condition for $u$ is $u=0$.

\subsubsection{Fermion-boson vertex}

The fermion-boson vertex has the form
\begin{eqnarray}
 \Gam_{\psi^2\phi^*} =
 g \int_{k\alf} \int_q \left(
  \psib_{-k,\alf\down} \psib_{k+q,\alf\up} \,
  \phi_q +
  \psi_{k+q,\alf\up} \psi_{-k,\alf\down} \,
  \phi^*_q \right) \, .
\end{eqnarray}
The fermion-boson vertex is not renormalized within our truncation. 
The usual one-loop vertex correction, which is formally of order $g^3$, 
vanishes in the normal phase due to particle conservation.
\cite{strack08}
Hence, the coupling $g$ remains invariant at its bare value $g=1$ 
in the course of the flow.
By contrast, the fermion-boson vertex obtained by decoupling the 
interaction of the Gross-Neveu model \cite{gross74} via a 
Hubbard-Stratonovich field is renormalized already at one-loop level.

Symmetry breaking in interacting Fermi systems is frequently studied
by extending the model to an arbitrary number of fermion flavors
$N_f$, and expanding in the parameter $1/N_f$. 
Our truncation captures the leading contributions for large $N_f$.
Below we will see that the low energy behavior is captured correctly 
also to leading order in $\eps$, where $\eps = 3-d$ is the
deviation from the critical spatial dimension $d_c=3$, below which 
anomalous scaling sets in.

\subsection{Flow equations}

In this subsection, we derive the flow equations for our truncated 
effective action. 
Both propagators, Eqs.~(\ref{eq:pp_green_f}, \ref{eq:pp_green_b}),
display singularities for vanishing momenta and frequencies, the 
zero-temperature fermion propagator everywhere in the semimetallic phase, 
and the boson propagator at the QCP when the bosonic mass vanishes. 
These infrared singularities are regularized by adding optimized 
momentum cutoffs \cite{litim01} for fermions (subscript f) and 
bosons (subscript b),
\begin{eqnarray}
 R_{f\alf}^{\Lambda}(\bk) &=& 
 Z_f \left(-\Lambda\,\text{sgn}\left[\xi_{\bk\alf}\right]
 + \xi_{\bk\alf}\right)
 \theta\left[\Lambda - |\xi_{\bk\alf}|\right] \nonumber \\
 R_b^{\Lambda}(\bq) &=& 
 Z_b \left( - \Lambda^2 + \bq^2 \right) \theta\left[\Lambda^2 -
 \bq^2\right] \; ,
\label{eq:cutoffs}
\end{eqnarray}
to the inverse of the propagators.
We denote the regularized propagators by $G_{f\alf}^R$ and $G_b^R$. 
Note that we have set $v_f = 1$, such that $\Lam$ is a common
momentum cutoff for fermions and bosons, which is a frequent
choice for critical fermion-boson theories.\cite{litim01,gies04} 
The scale-derivatives of the cutoffs read,
\begin{eqnarray}
 \partial_{\Lambda} R_{f\alf}^{\Lambda}(\bk) &=& 
 -Z_f \, \text{sgn}\left[\xi_{\bk\alf}\right] \, 
 \theta\left[\Lambda - |\xi_{\bk\alf}|\right] \nonumber\\
 \partial_{\Lambda} R_b^{\Lambda}(\bq) &=& 
 - 2 Z_b \Lambda \, \theta\left[\Lambda^2 - \bq^2 \right] \, ,
\end{eqnarray}
where terms proportional to $\partial_{\Lam} Z_f$ and 
$\partial_{\Lam}Z_b$ are neglected. 
These additional terms are of higher order in the vertices.
Arguments buttressing this commonly employed approximation
are given in Ref.~\onlinecite{berges02}.

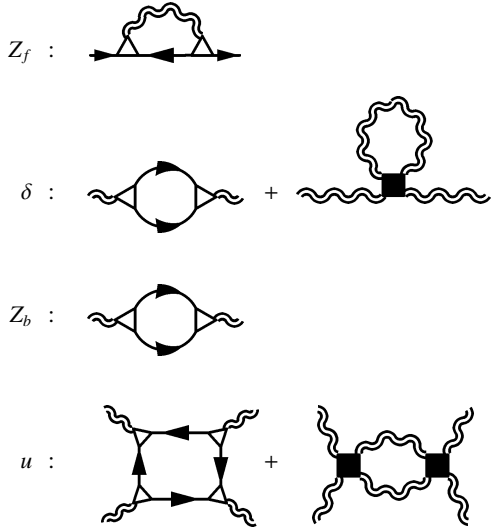
\begin{figure}
\begin{fmffile}{20081208_dirac_4}
\begin{eqnarray}
Z_{f}&:&
\parbox{25mm}{\unitlength=1mm\fmfframe(2,2)(1,1){
\begin{fmfgraph*}(20,20)\fmfpen{thin}
\fmfleft{l1}
 \fmfright{r1}
 \fmfpolyn{empty,tension=0.4}{G}{3}
 \fmfpolyn{empty,tension=0.4}{K}{3}
  \fmf{fermion}{l1,G1}
   \fmf{fermion,straight,tension=0.5,left=0.}{K3,G2}
  \fmf{dbl_wiggly,tension=0.,left=0.7}{G3,K2}
  \fmf{fermion}{K1,r1}
 \end{fmfgraph*}
}}\nonumber\\[-8mm]
\delta&:&
\parbox{25mm}{\unitlength=1mm\fmfframe(2,2)(1,1){
\begin{fmfgraph*}(20,20)\fmfpen{thin} 
 \fmfleft{l1}
 \fmfright{r1}
 \fmfpolyn{empty,tension=0.3}{G}{3}
 \fmfpolyn{empty,tension=0.3}{K}{3}
  \fmf{dbl_wiggly}{l1,G1}
 \fmf{fermion,tension=0.2,right=0.6}{G2,K3}
 \fmf{fermion,tension=0.2,left=0.6}{G3,K2}
 \fmf{dbl_wiggly}{K1,r1}
 \end{fmfgraph*}
}}+
\parbox{25mm}{\unitlength=1mm\fmfframe(2,2)(1,1){
\begin{fmfgraph*}(25,25)\fmfpen{thin} 
 \fmfleft{l1}
 \fmfright{r1}
 \fmftop{v1}
 \fmfpolyn{full}{G}{4}
 \fmf{dbl_wiggly,straight}{l1,G4}
 \fmf{dbl_wiggly,straight}{G1,r1}
 \fmffreeze
\fmf{dbl_wiggly,tension=0.1,right=0.7}{G2,v1}
\fmf{dbl_wiggly,tension=0.1,right=0.7}{v1,G3}
\end{fmfgraph*}
}}\nonumber\\[-8mm]
Z_{b}&:&
\parbox{25mm}{\unitlength=1mm\fmfframe(2,2)(1,1){
\begin{fmfgraph*}(20,15)\fmfpen{thin} 
 \fmfleft{l1}
 \fmfright{r1}
 \fmfpolyn{empty,tension=0.3}{G}{3}
 \fmfpolyn{empty,tension=0.3}{K}{3}
  \fmf{dbl_wiggly}{l1,G1}
 \fmf{fermion,tension=0.2,right=0.6}{G2,K3}
 \fmf{fermion,tension=0.2,left=0.6}{G3,K2}
 \fmf{dbl_wiggly}{K1,r1}
 \end{fmfgraph*}
}}\nonumber\\
u&:&
\parbox{25mm}{\unitlength=1mm\fmfframe(2,2)(1,1){
\begin{fmfgraph*}(25,15)
\fmfpen{thin}
\fmfleftn{l}{2}\fmfrightn{r}{2}
 \fmfpolyn{empty,tension=0.8}{OL}{3}
 \fmfpolyn{empty,tension=0.8}{OR}{3}
 \fmfpolyn{empty,tension=0.8}{UR}{3}
 \fmfpolyn{empty,tension=0.8}{UL}{3}
   \fmf{dbl_wiggly}{l1,OL1}
   \fmf{dbl_wiggly}{l2,UL1}
   \fmf{dbl_wiggly}{OR1,r1}
   \fmf{dbl_wiggly}{UR1,r2}
  \fmf{fermion,straight,tension=0.5}{OL2,OR3}
  \fmf{fermion,straight,tension=0.5}{UR3,OR2}
  \fmf{fermion,straight,tension=0.5}{UR2,UL3}
  \fmf{fermion,straight,tension=0.5}{OL3,UL2}
 \end{fmfgraph*}
}}+
\parbox{25mm}{\unitlength=1mm\fmfframe(2,2)(1,1){
\begin{fmfgraph*}(25,15)
\fmfpen{thin} 
\fmfleftn{l}{2}\fmfrightn{r}{2}
\fmfrpolyn{full}{G}{4}
\fmfpolyn{full}{K}{4}
\fmf{dbl_wiggly}{l1,G1}\fmf{dbl_wiggly}{l2,G2}
\fmf{dbl_wiggly}{K1,r1}\fmf{dbl_wiggly}{K2,r2}
\fmf{dbl_wiggly,left=.5,tension=.3}{G3,K3}
\fmf{dbl_wiggly,right=.5,tension=.3}{G4,K4}
\end{fmfgraph*}
}}\nonumber
\end{eqnarray}
\end{fmffile}
\caption{Feynman diagrams representing the flow equations.}
\label{fig:dirac_flow}
\end{figure}

The recipe to obtain the flow equations is now the following: 
one executes a cutoff-derivative acting on $R_{f,b}^{\Lambda}$ 
in the analytic expressions corresponding to all 1-loop 
one-particle-irreducible Feynman diagrams for the parameters 
$Z_{f}$, $Z_{b}$, $\delta$, $u$, which can be constructed with
the interaction vertices and the regularized propagators 
$G_{f\alf}^R$ and $G_b^R$, as shown in Fig.~\ref{fig:dirac_flow}. 
For the cutoff derivative and loop integration we use the
short-hand notation
\begin{eqnarray}
 \int_k^R = 
 \int \frac{d k_0}{2\pi} \int \frac{d^dk}{\left(2\pi\right)^{d}}
 \sum_{s=b,f} \left(-\dot{R}_s^{\Lambda}\right)
 \partial_{R_s^{\Lambda}} \; .
\end{eqnarray}

The feedback of bosonic fluctuations on the fermionic propagator
is captured by the self-energy diagram in the first line of 
Fig.~\ref{fig:dirac_flow}, leading to the flow equation for
the fermionic $Z$-factor
\begin{eqnarray}
\partial_{\Lambda} Z_f = 
 g^2 \int_q^R \left. \frac{\partial}{i\partial k_0} \,
 G_{f\alf}^R(q-k) \, G_b^R(q) \, \right|_{k=0} \; .
\label{eq:Zf}
\end{eqnarray}
Note that there is no sum over $\alf$ here, and $\alf=\pm 1$ 
yields the same $Z_f$.


For the control parameter, we evaluate the diagrams with two 
external bosonic legs, obtaining the two contributions
\begin{eqnarray}
 \partial_{\Lambda}\delta &=& 
 g^2 \int_{k\alf}^R G_{f\alf}^R(k) \, G_{f\alf}^R(-k) + 
 \frac{u}{2} \int_q^R G_b^R(q) \, .
\label{eq:delta}
\end{eqnarray}
The fermionic contribution on the right-hand-side is positive 
leading to a reduction of $\delta$ for decreasing $\Lambda$, 
whereas the bosonic contribution tends to increase $\delta$. 

The flow of the bosonic $Z$-factor is obtained as the second 
frequency derivative of the fermionic particle-particle bubble:
\begin{eqnarray}
\partial_{\Lambda}Z_{b} = 
 g^2 \int_{k\alf}^R \frac{1}{2} \left. 
 \frac{\partial^2}{\partial q_0^2} 
 G_{f\alf}^R(k+q) \, G_{f\alf}^R(-k) \, \right|_{q=0} \, .
\label{eq:Zb}
\end{eqnarray}
The bosonic tadpole diagram does not contribute here, 
as the $\phi^{4}$-vertex $u$  is taken as momentum- and 
frequency-independent.

Finally, the bosonic self-interaction flows according to
\begin{eqnarray}
 \partial_{\Lambda} u =
 &-& 4 g^4 \int_{k\alf}^R 
 \big[ G_{f\alf}^R(-k) \big]^2 \big[ G_{f\alf}^R(k) \big]^2 
 \nonumber\\
 &+& \frac{5}{4} u^2 \int_q^R \big[ G_b^R(q) \big]^2 \; ,
\label{eq:u}
\end{eqnarray}
where the first terms generates $u$ and the second, bosonic 
term tends to reduce $u$ in the course of the flow.

All frequency and momentum integrations in the above flow 
equations can be performed analytically. 
We now eliminate explicit $\Lam$-dependences from the flow
equations by employing the following scaling variables:
\begin{eqnarray}
 \tilde{\delta} &=& \frac{\delta}{\Lambda^{2}Z_{b}} \nonumber\\
 \tilde{g} &=&
 \frac{g\sqrt{K_{d}}}{\Lambda^{\frac{3-d}{2}}Z_{f}\sqrt{Z_{b}}\sqrt{d}}
 \nonumber\\
 \tilde{u} &=& \frac{u\,K_{d}}{\Lambda^{3-d}Z_{b}^{2}d}\,\,.
 \label{eq:def_variables}
\end{eqnarray}
Dependences on the $Z$-factors are absorbed by introducing
the anomalous dimensions for fermionic and bosonic fields,
\begin{eqnarray}
\eta_{f}&=&-\frac{d \log Z_{f}}{d \log \Lambda}\nonumber\\
\eta_{b}&=&-\frac{d \log Z_{b}}{d \log \Lambda} \; .
\label{eq:def_etas}
\end{eqnarray}
The flow equations for the control parameter and the bosonic 
self-interaction are then obtained as
\begin{eqnarray}
 \frac{d\tilde{\delta}}{d\log \Lambda} &=& 
 \left(\eta_b - 2 \right) \tilde{\delta}
 + \tilde{g}^2 
 - \frac{\tilde{u}}{4\left(1+\tilde{\delta}\right)^{3/2}}
 \nonumber\\[3mm]
 \frac{d\tilde{u}}{d\log \Lambda} &=&
 \left(d-3+2\eta_b \right)\tilde{u} - 6\, \tilde{g}^4
 + \frac{15}{16}\, \frac{\tilde{u}^2}
 {\left(1+\tilde{\delta}\right)^{5/2}} \; . \nonumber\\
\label{eq:betas_qcp}
\end{eqnarray}
The rescaled fermion-boson vertex $\tilde{g}$ obeys the equation:
\begin{eqnarray}
 \frac{d\tilde{g}}{d\log \Lambda} &=&
 \left(\eta_{f} +\frac{1}{2} \eta_{b} - 
 \frac{3-d}{2}\right)\tilde{g} \,\, .
\label{eq:beta_g}
\end{eqnarray}
The equations (\ref{eq:betas_qcp}, \ref{eq:beta_g}) have to be 
considered in conjunction with the boson and fermion anomalous 
dimensions:
\begin{eqnarray}
 \eta_b &=& \frac{3}{4}\tilde{g}^{2} \nonumber\\
 \eta_f &=& \frac{\tilde{g}^2}{2}
 \left(
 \frac{1}{\left(1+\tilde{\delta}\right)^{3/2}} +
 \frac{2}{\left(1+\tilde{\delta}\right)}\right)
 \frac{1}
 {2+\tilde{\delta}+2\sqrt{1+\tilde{\delta}}} \; .
 \label{eq:dirac_etas}
\end{eqnarray}
We recall that we have set the Fermi velocity $v_f = 1$. 
We will now investigate the fixed point of these equations 
and then solve them numerically.

 
\section{Solution at the quantum critical point}
\label{sec:sol_qcp}

\subsection{Fixed point}

\begin{figure}[b]
\begin{center}
\vspace*{3mm}
\includegraphics*[width=72mm]{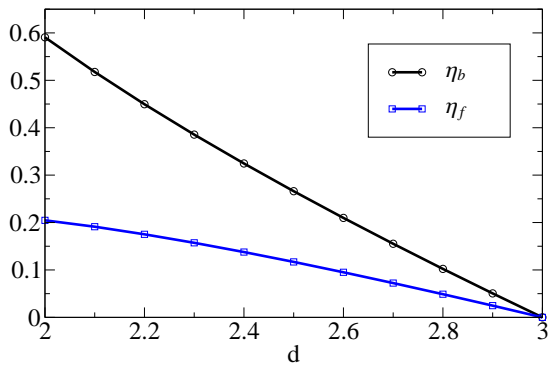}
\caption{(Color online)
 Fermion and boson anomalous exponents at the QCP for 
 $2\leq d \leq 3$.}
\label{fig:fp_line}
\end{center}
\end{figure}

At the quantum critical point, the system is scale-invariant
at low energies. The scaling regime is associated with a fixed
point of the flow equations, which attracts the flow for small 
$\Lambda$. The fixed point is defined by the condition that the
right hand sides of the flow equations for $\tilde\delta$,
$\tilde u$, and $\tilde g$ vanish, such that
\begin{eqnarray}
 \frac{d\tilde{\delta}}{d\log \Lambda} =
 \frac{d\tilde{g}}{d\log \Lambda} = 
 \frac{d\tilde{u}}{d\log \Lambda}=0 \, .
\label{eq:constraint}
\end{eqnarray}
One thus has to solve three algebraic equations together with the 
equations for the anomalous exponents, Eq.~(\ref{eq:dirac_etas}).
The fermion-boson vertex $\tilde{g}$ and the boson self-interaction 
$\tilde{u}$ are relevant couplings below three dimensions,
and the flow equations for $\tilde{g}$ and $\tilde{u}$ have 
stable non-Gaussian $\left(\tilde{g}\neq 0,\; \tilde{u}\neq 0\right)$ 
solutions with finite anomalous exponents $\eta_f$ and $\eta_b$.
The condition $\frac{d\tilde g}{d\log\Lambda} = 0$ directly 
implies a relation between the anomalous dimensions:
\begin{equation}
 \eta_f + \frac{\eta_b}{2} = \frac{3-d}{2} \; .
\label{eq:inter_etas}
\end{equation}
The values of $\eta_f$ and $\eta_b$ as obtained from a numerical
solution of the fixed point equations are plotted in 
Fig.~\ref{fig:fp_line} as a function of dimensionality $d$
for $2 \leq d \leq 3$.

The anomalous dimensions vanish at the critical dimension
$d_c = 3$. It is instructive to solve the fixed point equations
analytically to leading order in $\eps = 3 - d$.
For the anomalous dimensions one finds
\begin{eqnarray}
 \eta_f &=& \eps/4 \; , \nonumber \\
 \eta_b &=& \eps/2 \; .
\end{eqnarray}
The fixed point values for $\tilde g^2$, $\tilde u$ and 
$\tilde\delta$ are all of order $\eps$.
Terms neglected in our truncation are therefore of higher order 
in $\eps$. 
Note that the above results for $\eta_f$ and $\eta_b$ differ 
from the anomalous dimensions obtained for the Gross-Neveu model 
\cite{vasilev93,karkkainen94,zinnjustin96} already to leading 
order in $\eps$.

At the quantum critical point, the fermionic properties of the 
system cannot be described in terms of Fermi liquid theory. 
A finite fermion anomalous dimension entails a fermion propagator 
scaling with an anomalous power law as
\begin{eqnarray}
 G_{f\alf}(\lambda\bk,\lambda\omega) \propto \lambda^{\eta_f - 1} 
\label{eq:pp_qcp_etas}
\end{eqnarray}
for $\lam \to 0$, corresponding to a non-analytic frequency and
momentum dependence of the self-energy. In particular,
\begin{eqnarray}
 \Sigma_{f\alf}(\b0,\omega) \sim \omega^{1-\eta_{f}} \;
\label{eq:ferm_self_qcp}
\end{eqnarray}
at the Fermi point $\bk=\b0$.
In two dimensions we obtain $1-\eta_f=0.80$. 
A breakdown of Fermi liquid theory in quantum critical interacting
Fermi systems is very common.\cite{loehneysen07}
For example, at a quantum critical point associated with a d-wave
Pomeranchuk instability in two dimensions, the fermionic self-energy
scales as $\omega^{2/3}$.\cite{metzner03,dellanna06} 
The fermion propagator also develops a small anomalous dimension in 
QED$_{3}$ \cite{franz02} and in the Gross-Neveu model.
\cite{rosa01,vasilev93,karkkainen94,zinnjustin96}

Concerning the collective properties of the system, a finite 
boson anomalous dimension leads to anomalous scaling of the 
order parameter propagator 
\begin{eqnarray}
 G_b(\lambda\bq,\lambda\omega) \propto \lambda^{\eta_b - 2}
\label{eq:pp_qcp_bosonprop}
\end{eqnarray}
for $\lambda \to 0$, where $2-\eta_b=1.41$ in two dimensions.

In a different context, at the antiferromagnetic QCP of the
spin-fermion model in two dimensions, Abanov \textit{et al.} 
\cite{abanov00,abanov03,abanov04} obtained an anomalous momentum
scaling $\propto |\bq|^{-1.75}$ for the spin susceptibility within 
a perturbative $1/N$ calculation, where $N$ is the number of hot 
spots on the Fermi surface.

Anomalous scaling of Dirac-like fermions and order parameter 
fields has also been found previously for certain quantum phase
transitions in d-wave superconductors, where a second order 
parameter forms on top of d-wave superconductivity.
\cite{vojta00,huh08}

\subsection{Quantum critical flows in two dimensions}
\label{subsec:qcflows}

\begin{figure}
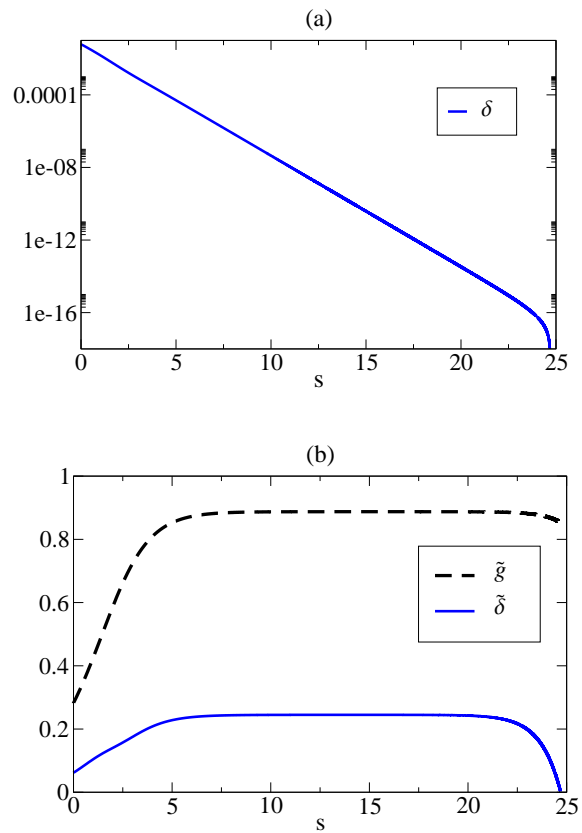

\includegraphics*[width=74mm]{dong_delta_bare_cone.eps}\\[7mm]
\hspace*{4mm}
\includegraphics*[width=72mm]{dong_del_g_cone.eps}
\caption{(Color online)
 (a): Flow of the control parameter $\delta$ at the QCP.
 (b): Flows of the rescaled control parameter and fermion-boson 
 vertex as a function of 
 $s = - \ln\left[\Lambda/\Lambda_{0}\right]$. 
 The ultraviolet (infrared) regime 
 is on the left (right) side of the plots.}
\label{fig:delta_cone_flows}
\vspace{0mm}
\end{figure}

\begin{figure}
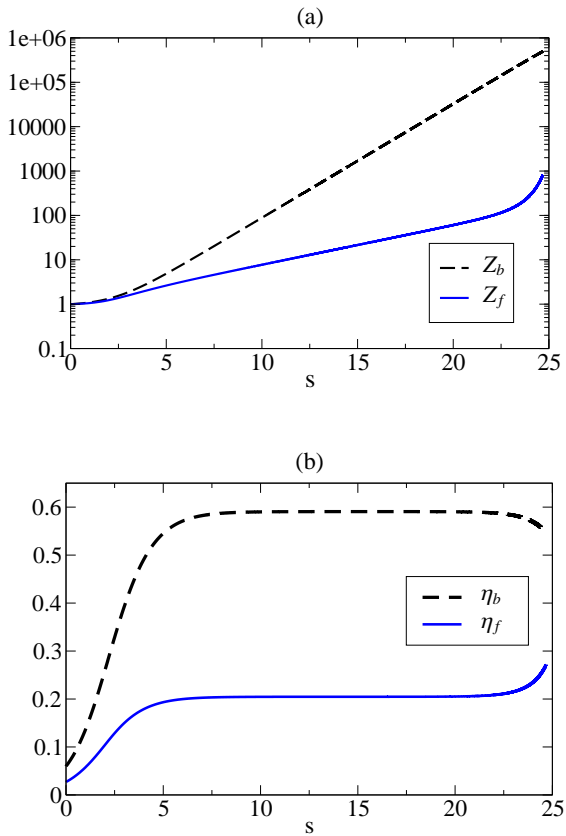

\includegraphics*[width=74mm]{dong_Zflows.eps}\\[8mm]
\hspace*{3mm}
\includegraphics*[width=71mm]{dong_etas_flows_cone.eps}
\caption{(Color online)
 (a): Flows of the bosonic and fermionic frequency- and momentum factors 
 $Z_b$ and $Z_f$.
 (b): Scaling plateaus of the fermion and boson anomalous dimensions at 
 quantum criticality.}
\label{fig:eta_cone_flows}
\vspace{0mm}
\end{figure}

We now establish a continuous link between the bare action and 
the low energy behavior of the \emph{effective}\/ action at the 
QCP in two dimensions. 
To this end, we compute the renormalization group flow as a 
function of $\Lam$ by solving the flow equations 
(\ref{eq:betas_qcp}-\ref{eq:dirac_etas}) numerically.
The initial conditions of our parameters are chosen to match 
the bare action in Eq.~(\ref{eq:dirac_finalmodel}), regularized
by an additional term $\int \phi_q^* (q_0^2 + \bq^2) \, \phi_q$
as discussed in Sec.~IVA2: $u=0$, $Z_b = Z_f = g = 1$. 
The Fermi velocity and the ultraviolet cutoff are set to unity,
$\Lambda_0 = v_f =1$. 
To reach the quantum critical point, the initial value of $\delta$ 
is tuned such that $\delta \to 0$ at the end of the flow. 
The critical initial value for $\delta$ obtained in this way is
$\delta_c = 0.0616$.
This is smaller than the mean-field value 
$\delta_c^{\rm MFT} = \frac{1}{2\pi}$ obtained in Sec.~III for
two distinct reasons.
One is a simple offset in the initial condition for the flow, 
since the regulator functions, Eq.~(\ref{eq:cutoffs}), are not 
infinite for $\Lam = \Lam_0$.
Running a ''mean-field flow'' with these regulators but 
discarding bosonic fluctuations would yield
$\delta_c^{\rm MFT*} = \frac{1}{4\pi} \approx 0.0796$.
The remaining reduction is due to fluctuations.

\begin{figure}
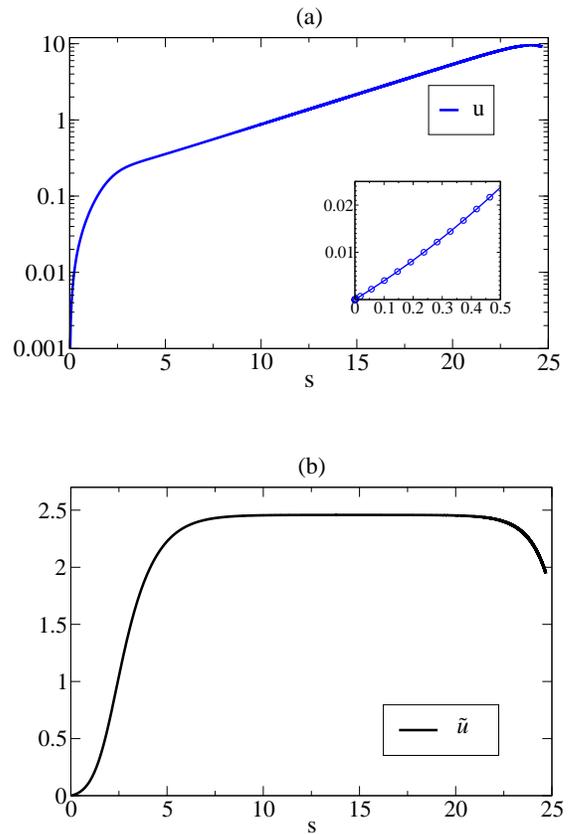

\includegraphics*[width=73mm]{dong_u_flow.eps}\\[-28mm]
\hspace*{34mm}
\includegraphics*[width=25mm]{dong_u_bare_inset.eps}\\[18.5mm]
\hspace*{3mm}
\includegraphics*[width=70mm,angle=0]{dong_u_cone.eps}\\[0mm]
\caption{(Color online)
 (a): Flow of the quartic coupling $u$. 
 Inset: Small-$s$ behavior where fermionic fluctuations start
 generating $u$.
 (b): Fixed point plateau of the rescaled $\phi^4$-coupling.}
\label{fig:u_cone_flows}
\vspace{0mm}
\end{figure}

In Fig.~\ref{fig:delta_cone_flows} (a), we show the flow 
of the control parameter at the QCP versus cutoff scale 
in a double logarithmic plot. 
We observe scaling behavior 
\begin{eqnarray}
 \delta \sim \Lambda^{2-\eta_b}
\end{eqnarray}
over approximately 15 (!) orders of magnitude (note the small 
values of $\delta$ on the vertical axis).
Scaling stops at large $s$ due to numerical errors. 
The power law scaling of $\delta$ corresponds to the fixed point 
plateaus of the \emph{rescaled} control parameter $\tilde{\delta}$ 
in Fig.~\ref{fig:delta_cone_flows}~(b), where also the flow of the
rescaled fermion-boson vertex is shown. 
The point we make here is that, starting from the microscopic model,
the effective action is really attracted by the quantum critical 
fixed point.

In Fig.~\ref{fig:eta_cone_flows} we present flows for the fermionic 
and bosonic $Z$-factors $Z_f$ and $Z_b$ as well as the associated 
anomalous dimensions $\eta_f$ and $\eta_b$. 
In Fig.~\ref{fig:eta_cone_flows} (a), we observe that $Z_{b}$
diverges as a power law: 
$Z_b \sim \Lambda^{-\eta_b}$, which corresponds to fixed point 
plateaus of $\eta_b$ shown in Fig.~\ref{fig:eta_cone_flows} (b). 
The fermionic $Z_f$ also diverges as a power law,
$Z_f \sim \Lambda^{-\eta_f}$, 
with a smaller slope than $Z_b$, as reflected by the lower fixed 
point plateau of $\eta_f$ in Fig.~\ref{fig:eta_cone_flows} (b). 
The numerical solution for the $Z$-factors fulfils the relation 
Eq.~(\ref{eq:inter_etas}) and the fixed point values match 
those of Fig.~\ref{fig:fp_line} at $d=2$.

We show flows of the $\phi^4$-coupling in Fig.~\ref{fig:u_cone_flows}. 
Fermion fluctuations quickly generate $u$, and then the interplay 
of fermionic and bosonic fluctuations leads to the power law scaling 
behavior
\begin{eqnarray}
 u \sim \Lambda^{3-d-2\eta_b} \; ,
\end{eqnarray}
accompanied by the fixed point plateaus of $\tilde{u}$ in 
Fig.~\ref{fig:u_cone_flows} (b).

\subsection{Quantum critical exponents}
\label{subsec:suszeb}

\begin{figure}
\begin{center}
\vspace*{1.5mm}
\includegraphics*[width=58mm]{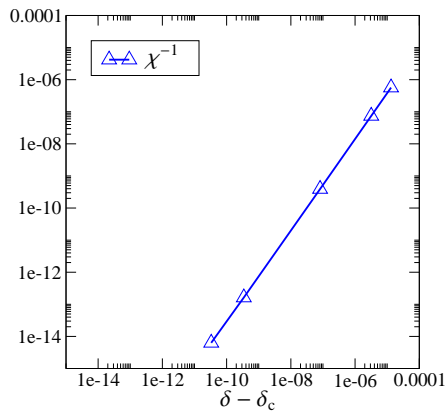}
\end{center}
\vspace*{-3mm}
\caption{(Color online)
Logarithmic plot of the inverse susceptibility versus 
 $\delta - \delta_c$ in $d=2$.}
\label{fig:dirac_nu}
\end{figure}

When approaching the quantum critical point along the control 
parameter axis, the susceptibility and the correlation length 
diverge as a power law,
\begin{eqnarray}
 \chi &\sim& (\delta - \delta_c)^{-\gamma} \; , \nonumber \\
 \xi  &\sim& (\delta - \delta_c)^{-\nu} \; ,
\end{eqnarray} 
where $\delta$ is the bare bosonic mass ($=-1/U$). 
The inverse susceptibility is given by the renormalized bosonic 
mass at the end of the flow, $\chi^{-1} = \lim_{\Lambda \to 0} \delta$.
Reading off the slope of a double-logarithmic plot of the susceptibility
as a function of $\delta - \delta_c$, see Fig.~\ref{fig:dirac_nu}, 
we find in two dimensions:
\begin{eqnarray}
 \gamma = 1.4 \; .
\end{eqnarray}
The correlation length exponent $\nu$ now follows from the general 
scaling relation $\gamma=\nu\left(2-\eta_{b}\right)$.\cite{goldenfeld92} 
With $\eta_b = 0.59$ in two dimensions, one obtains 
\begin{eqnarray}
 \nu = 1.0 \; .
\end{eqnarray}

\section{Conclusion}
\label{sec:pp_conclusion}

We have analyzed the quantum phase transition between a semimetal 
and a superfluid in a model of attractively interacting fermions
with a Dirac cone dispersion. 
The model is a prototype for systems for which the Hertz-Millis approach 
to quantum phase transitions in itinerant Fermi systems is not applicable,
since integrating out the fermions leads to a singular Landau-Ginzburg
functional.
Using the functional RG framework, we have derived coupled flow equations 
for the gapless fermionic degrees of freedom and the bosonic order 
parameter fluctuations.
Fermions and bosons acquire anomalous dimensions at the QCP 
in dimensions $d < 3$. Consequently, both 
the fermion and the order parameter propagators 
are non-analytic functions of frequency and momentum. 
In two dimensions, the fermionic self-energy scales as $\omega^{0.8}$
at low frequencies, implying the absence of fermionic quasi-particles
and non-Fermi liquid behavior.
We have also computed the susceptibility and the correlation length 
exponents when approaching the QCP along the control parameter axis.

The strong impact of the gapless fermions on the order parameter
interaction spoils the Hertz-Millis approach to quantum phase 
transitions, which is based on the Landau-Ginzburg paradigm.
The fermionic anomalous dimension $\eta_f$ influences critical
exponents in a way that resembles the interplay of fermions and
bosons in the Gross-Neveu model.\cite{rosa01,zinnjustin96}
Recently, the Gross-Neveu model has been proposed as an effective
field theory describing the quantum phase transition between a
semimetal and an antiferromagnetic insulator for {\em repulsively}
interacting electrons on the two-dimensional honeycomb lattice.
\cite{herbut06}

In the present work we have focussed on ground state properties, 
but an extension to finite temperatures should be straightforward. 
In particular, the quantum critical regime above the QCP in the
$(\delta,T)$-plane could be studied with the same truncation of the
functional RG hierarchy.
Another interesting extension would be the inclusion of the pseudo
spin structure necessary to describe Dirac fermions on the honeycomb 
lattice.

Our work may also serve as a guideline for the study of more complex
quantum phase transitions in itinerant systems, for which the 
Hertz-Millis approach is not applicable, or at least questionable, 
such as certain magnetic transitions in low dimensions.


\begin{acknowledgments}
We thank P.~Jakubczyk, V.~Juri\v{c}i\'c, B.~Obert, and O.~Vafek 
for valuable discussions, and P.~Jakubczyk also for helpful comments 
on the manuscript.
This work has been supported by the DFG research group FOR 723.
\end{acknowledgments}


\end{document}